\input epsf
\documentstyle[11pt]{article}

\advance\oddsidemargin by 8mm

\def\be{\begin{equation}}
\def\ee{\end{equation}}

\def\half{\frac{1}{2}}
\def\third{\frac{1}{3}}
\def\quart{\frac{1}{4}}

\def\ph3{$\phi^3$}
\def\gst{\gamma_{str}}
\def\gstwo{\gamma^{I \! I}}

\def\mgap{\qquad \qquad}

\def\ltsim{\lower3pt\hbox{$\, \buildrel < \over \sim \, $}}
\def\gtsim{\lower3pt\hbox{$\, \buildrel > \over \sim \, $}}

\def\gone{{{\cal G}_r^I}}
\def\zone{{\cal Z}^{I}}
\def\xcone{{x_c^{I}}}
\def\mone{M^{I}}

\def\gtwo{{{\cal G}_r^{I\!I}}}
\def\gmark{{{\cal G}_2^I(r)}}
\def\ztwo{{\cal Z}^{I\!I}}
\def\xctwo{{x_c^{I\!I}}}
\def\mtwo{{M^{I\!I}}}

\def\nmean{{\langle n \rangle}}
\def\Nmean{{\langle N \rangle}}
\def\smean{{\langle \Sigma S \rangle}}

\def\fbar{\bar{f}}
\def\hbar{\bar{h}}

\def\mce{{\cal M}_{ce}}

\def\lcal{{\cal L}}
\def\setloop{\lcal_G}
\def\gover{\overline{G}}

\def\xld{(x',\{\lambda'\})}

\begin{document}

\vbox{\smash{\vbox{
\begin{flushright}
 \large NBI-HE-96-04 \\
 \large January 31, 1996
\end{flushright}}}
\title{Correlation Functions in the Multiple Ising Model
 Coupled to Gravity} 
\author{M. G. HARRIS and J. AMBJ{\O}RN \\ \\
  \small Niels Bohr Institute, Blegdamsvej 17, \\
  \small DK-2100 Copenhagen, Denmark. \\ 
  \small E-mail addresses: Martin.Harris@nbi.dk, ambjorn@nbi.dk}
\date{}
\maketitle}
\begin{abstract}
The model of $p$ Ising spins coupled to 2d gravity, in the form of
a sum over planar \ph3 graphs, is studied and in particular the two-point
and spin-spin correlation functions are considered. We first
solve a toy model in which only a partial summation over spin
configurations is performed and, using a modified geodesic distance,
various correlation functions are determined. The two-point
function has a diverging length scale associated with
it. The critical exponents are calculated and it is shown that all
the standard scaling relations apply. Next the full model is studied,
in which all spin configurations are included. Many of the
considerations for the toy model apply for the full model, which also
has a diverging geometric correlation length associated with the
transition to a branched polymer phase. Using a transfer function we
show that the two-point and spin-spin correlation functions decay
exponentially with distance. Finally, by assuming various scaling
relations, we make a prediction for the
critical exponents at the transition between the magnetized and
branched polymer phases in the full model. 

\end{abstract}


\section{Introduction}
The work of Knizhnik, Polyakov and Zamolodchikov (KPZ)~\cite{KPZ}
as well as of David, Distler and Kawai (DDK)~\cite{DDK}
made it possible to understand many aspects
two-dimensional quantum gravity coupled to conformal field theories.
At the same time, it became clear that the models
of dynamical triangulations coupled to matter fields provide us with
a statistical mechanical realisation of the models described by KPZ and
DDK, in the same way as the two-dimensional Ising model at its critical
point can be viewed as a conformal $c=1/2$ field theory. Two aspects
of 2d quantum gravity coupled to matter fields remain
puzzling: the $c=1$ barrier and the concept of correlation length.
The formulae for critical exponents derived by KPZ cease to be valid
for $c >1$ and the critical exponents are derived by general scaling
arguments applied to globally defined operators. At no point is the
concept of a divergent correlation length introduced. For ordinary 
statistical systems the divergence of a correlation length
when a critical temperature is approached is believed to be the
underlying reason that general scaling arguments work well.
The difficulties of defining a local length scale in quantum gravity are
well known and have so far prevented a proper treatment of correlation
functions by means of continuum methods.

Working entirely in the context of dynamical
triangulations the problems mentioned are not seen directly.
Statistical models with multiple Ising spins living on
dynamically triangulated surfaces are perfectly
well defined even if $c >1$ and they have a critical point.
At least superficially, the only difference compared to
$c <1$ is that we cannot solve the theory.
Nevertheless, low temperature expansions and mean field calculations seem
reliable for $c \to \infty$, as is confirmed by the agreement
between the theoretical calculations and Monte Carlo simulations of the
systems with large $c$~\cite{MC,Wexlerq}. 
The picture which emerges for large $c$
from the low temperature expansion~\cite{Wexler} is as follows:
for large $\beta$ (low temperature) there is a magnetized phase
for which $\gamma_{str}=-1/2$ separated from a branched polymer phase,
where $\gamma_{str}=1/2$, by a transition at a critical $\beta^*$ where
$\gamma^*_{str}=1/3$.

Likewise it has been possible by the use of dynamical triangulations to
address the question of correlation length in two-dimensional
quantum gravity. A two-point correlator between ``punctures''
has been calculated in pure gravity as a function of a ``quantum''
geodesic distance and it is found that standard scaling relations,
known from the theory of critical phenomena, are satisfied, although
with unusual critical exponents~\cite{AmbWat95}. 
So far it has not been possible in 2d quantum gravity
to calculate correlation functions of matter fields
as a function of the geodesic distance. The attempts to measure
the spin-spin correlation functions by Monte Carlo
simulations and to define a divergent
correlation length as one approaches the critical point, have so far
been ambiguous~\cite{CTBJ,MC}. 
The question arises as to whether there is a divergent
spin-spin correlation length associated with the phase transition between
a magnetized and a non-magnetized phase in the two-dimensional Ising
model coupled to gravity.
Rather surprisingly from a continuum point of view, we can,
using dynamical triangulations, begin to answer this question
in the limit of large $c$, i.e. in the limit where
a large number of Ising models is coupled to quantum gravity.

The statistical model we will define and solve
in the following sections is a toy model of 2d quantum gravity
coupled to Ising spins in the sense that the correct
summation over all triangulations is performed,
but not all the spin configurations
are included. The spin configurations we include are precisely the
ones which dominate in the large $c$ limit, at large $\beta$, namely
those for which the domains are connected in a tree-like fashion with
domain boundaries of minimal size. This model
allows us to calculate the two-point functions (using a certain
definition of distance, which is based on the geodesic distance)
and extract the critical
exponents. We will verify that these exponents satisfy
standard scaling relations and that the geometrical interpretation of
some of the exponents is related to the fractal structure of the underlying
``space-time'', in agreement with general arguments; that is, we
explicitly verify that the exponent $\nu$ is related to the Hausdorff
dimension by $\nu=1/d_H$.
The model has a third order transition between a tree-like (i.e.\ branched
polymer) phase and a magnetized phase. It is found that there is a
diverging correlation length in the tree phase associated with the
geometric two-point function, but no diverging correlation length
associated with the spin-spin correlation function.

Next we turn our attention to the full model of $p$ independent Ising
spins coupled to 2d gravity, which has a central charge of $c=p/2$. 
Following the 
analysis of the toy model, we show that various two-point functions
decay exponentially with distance in the magnetized phase. Again there
is a diverging correlation length associated with the geometric
two-point function. By assuming that all the standard scaling
relations still hold for this model and making a few further fairly
modest assumptions, we show that the critical exponents for the
transition between the magnetized and branched polymer phases, in the
full model, are the same as those in the toy model.

The rest of this article is organized as follows: in
section \ref{sec:toynomag} we
define the toy model, define a variant of geodesic distance in the model and
calculate the two-point function. In sections~\ref{sec:toycrit}
and~\ref{sec:dh} 
the critical exponents and Hausdorff dimension are calculated for the
toy model, whilst 
section \ref{sec:spinspin} discusses the spin-spin correlation function.
In section \ref{sec:toymag}
we verify by direct calculation in an external magnetic
field, that the magnetic exponents found by scaling arguments are indeed
the correct ones. In section \ref{sec:full}
we address some of the questions mentioned
above for the full model, i.e.\ in the model where
we sum over all the spin configurations. Finally, section \ref{sec:concl}
contains our conclusions.

\section{Toy model, without a magnetic field}
\label{sec:toynomag}
\subsection{Definition}

Before looking at the full model of $p$ spins coupled to 2d gravity
(in the form of a sum over planar \ph3 graphs), we will solve a toy
model, which is very similar to the one studied in~\cite{ADJ}.
As for that model we will show that there is a magnetized phase for
which $\gst=-\half$ and a branched polymer phase with $\gst=\half$. On
the boundary $\gst^*=\third$, which is the same value as that
in~\cite{ADJ,JonWhe94} 
and agrees with the result
$\gst^*=\overline{\gamma}/(\overline{\gamma}-1)$ for
$\overline{\gamma}=-\half$, the pure gravity value, which one might
expect from the calculation in~\cite{Dur94}.

\begin{figure}[bth]
\caption[l]{Measuring distances}
\label{fig:tree}
\begin{picture}(100,55)(0,0)
\centerline{\epsfbox{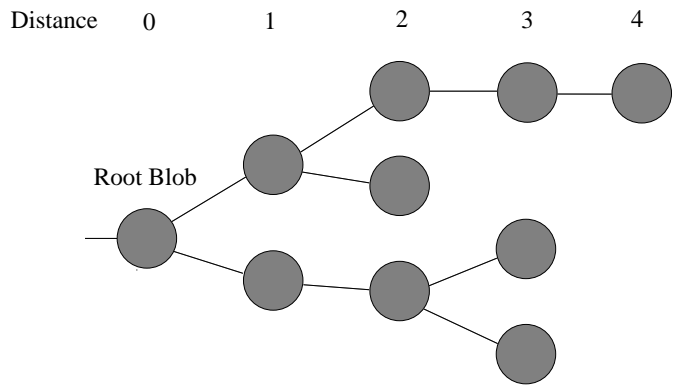}}
\end{picture}
\end{figure}

For our model we sum over all possible rooted planar \ph3 graphs
(let us denote this set of graphs by $\gone$), but we only sum over
a subset of the possible spin configurations, namely those for which
the domains are connected in a tree-like fashion, with at most one link
connecting any two domains. In this paper we shall only consider
planar graphs, that is, $\chi=2$ throughout. 
Each vertex will be weighted with a
factor of $x$ and has $p$ independent spins on it which can take the
values $\pm 1$ (i.e.\ vertex $i$ has spins $S^\alpha_i$ on it, with
$\alpha = 1, \cdots, p$). Links joining vertices with dissimilar spin
configurations will be given a factor of $e^{-2 \beta}$ for each spin
flavour which differs. Thus the grand canonical partition function, T, is
\be
\label{eqn:gcpf}
 T = \sum_{G \in \gone} x^N \sum_{\{S\}'} \prod_{<i j>} \exp
\left( \sum_{\alpha=1}^{p} \beta \left( \{S_i^\alpha S_j^\alpha -1
\right) \right) ,
\ee
where the first summation is over graphs, with $N$ being the number of
vertices in the graph. The product is over the
nearest neighbour pairs on the graph $G$ (referred to as ``links'') 
and the second summation is over
the following set of spin configurations. Take the graph $G$ and
decompose it into a set of one-particle irreducible (``1PI'') graphs
(which we shall call ``blobs'') connected in a tree-like fashion (see
fig.~\ref{fig:tree}); note that these blobs are essentially just
minimum necked baby universes (``minbu''s). 
The blobs are fully magnetized, that is, all the
vertices within a 1PI blob have the same set of spins, and we will sum
over the $2^p$ possible spin configurations for each blob (except 
the root blob for which all the spins will be fixed to be $+1$).

In the next sections we will solve this model exactly and also
calculate various correlation functions, but in order to do this we
need first to define how distances are to be measured on the graphs.
One possible measure of the distance between two vertices $A$ and $B$, is
the geodesic distance between them. That is, find the shortest path
from $A$ to $B$ along links, counting a distance of one unit for every
link traversed---this gives the geodesic distance between the two points. 
Unfortunately, this is quite difficult to deal with
analytically, although some results are known for the pure gravity
case~\cite{KKMW93}. In this paper we use a slightly different definition
of distance, which is much simpler to handle. The distance between two
vertices will be defined as the shortest path between them along
links, counting a distance of one unit only for links that separate
two 1PI blobs.

Since we are interested in correlations between the spins in the root
blob, which are held fixed and spins further away, distances will be
measured from the root blob.
Thus all the vertices in the root blob have a distance zero, all the
vertices within blobs connected by a single link to the root blob are
at distance one and so on (see fig.~\ref{fig:tree}). It will turn out
that the average number of vertices in each blob is very small, so
that for many graphs this definition of distance will be quite similar
to the geodesic distance, especially in the branched polymer phase for
which the partition function is dominated by tree-like graphs.

The advantage of measuring distances in this way is that it makes it very
easy to define a transfer function, $f(y)$,
\be
\label{eqn:f}
 f(y)= \sqrt{1- \lambda y} \ \ztwo \left( x (1- \lambda
y)^{-\frac{3}{2}} \right) + \frac{\lambda^2}{4 x} y^2,
\ee
where $\lambda = 2 x \left(1+ e^{-2 \beta} \right)^p$ and
\be
\ztwo(x) = \sum_{G \in \gtwo } x^N .
\ee
This summation is over the set $\gtwo$ of rooted planar 1PI graphs.

This function, $f(y)$, takes a rooted 1PI
blob and glues an arbitrary number of trees, each weighted with
a factor of $y$, on to the blob. Note that each time we glue a tree on to a
link in the blob, we pick up a factor of $x$, for the new vertex that
is created, a factor of two because we can hang the tree in one of two
directions and a factor of $(1+e^{\-2 \beta})^p$ to take account of
the $2^p$ different ways the spins on the blob and those on the tree's
root blob can differ---this accounts for the factor of $\lambda$
multiplying each $y$.
When we sum over all possible rooted 1PI blobs
and all possible ways of gluing trees to links this gives the first
term in (\ref{eqn:f}). More specifically, the blob without any trees attached
has a weight of $\ztwo(x)$. Consider a rooted graph with $N$ vertices, it has
$L =\half (3N-1)$ internal links and we can add an arbitrary number of
trees to each link, giving a total contribution of $x^N (1 + \lambda y
+ (\lambda y)^2 + \cdots )^L$ for the graph. When we sum over all 1PI
graphs this gives a contribution of
\be
\sum_{G \in \gtwo} x^N (1- \lambda y)^{-L} = \sqrt{1 - \lambda y} \ 
\ztwo \left( x (1- \lambda y)^{-\frac{3}{2}} \right) .
\ee

The second term in (\ref{eqn:f}) comes from the special case in
which the root blob only consists of a single vertex, in which case
we get a factor of $x$ for the vertex, $(1+ e^{-2 \beta})^{2p}$ for
the possible ways the spins may differ and $y^2$ for the two trees.

\begin{figure}[thb]
\caption[l]{$T=f(T)$}
\label{fig:tft}
\begin{picture}(100,25)(0,0)
\centerline{\epsfbox{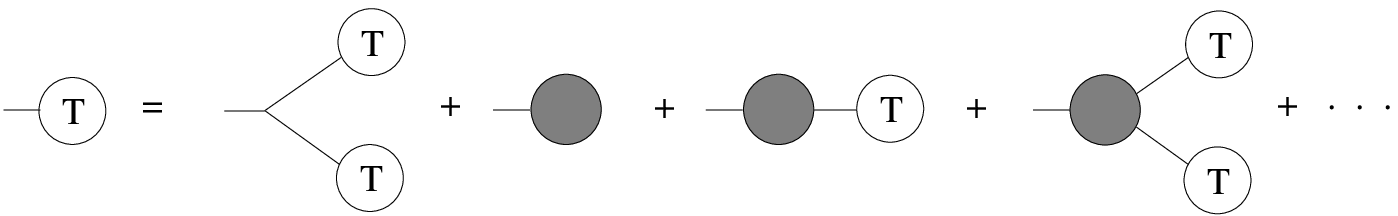}}
\end{picture}
\end{figure}

The partition function, $T$, satisfies $T=f(T)$ and this is represented
diagrammatically in fig.~\ref{fig:tft}; the shaded circle represents
the sum over 1PI graphs, $\ztwo(x)$. 
For fixed $p$ and $\beta$, this
equation gives $T$ as a function of $x$. By finding the closest
singularity to the origin, which occurs at $x_c(\beta,p)$, say, we can
determine the free energy for the model.
For what follows it will be convenient to define
$x' = x(1- \lambda T)^{-\frac{3}{2}}$.
Now, there are essentially two different types of singularity. One occurs when
$x'$ equals $\xctwo$, the critical value of $x'$ for $\ztwo(x')$ and
this will correspond to the magnetized phase. The other occurs when
the graphs become tree-like and we shall come back to this case later.

\subsection{$x_c$ in the magnetized phase}
Consider the first case, for which $x'=\xctwo$ and hence
$\ztwo(x')=\ztwo_c$, then writing $T(x_c)$ as $T_c$,
\be
T_c = \sqrt{1 - \lambda_c T_c} \ \ztwo_c + \frac{\lambda_c^2}{4 x_c}
T_c^2
\ee
and $\xctwo = x_c (1- \lambda_c T_c)^{-\frac{3}{2}}$, where $\lambda_c
= 2 x_c (1+e^{-2 \beta})^p$.
Defining,
\be
h \equiv \left(\frac{x_c}{\xctwo}\right)^\third = \sqrt{1 - \lambda_c
T_c} \ ,
\ee
we get
\be
\label{eqn:quadhh}
1- h^2 = \left(1 + e^{-2 \beta}\right)^p \left[   2 \xctwo \ztwo_c h^4
+ \half \left( 1 - h^2 \right)^2 \right].
\ee
However, $\ztwo(x)$ is a known function, from Br\'ezin {\it et al}
\cite{BIPZ}
one can show that
\be
\label{eqn:model2z}
\ztwo(x)= \frac{1}{x} \tau (1-3 \tau),
\ee
where $x^2= \tau (1-2 \tau)^2$.
This gives $\tau_c =\frac{1}{6}$, $\xctwo=\sqrt{\frac{2}{27}}$ and
$\ztwo_c = \frac{1}{4}\sqrt{\frac{3}{2}}$.
Hence we can solve (\ref{eqn:quadhh}) to get $x_c=\xctwo h^3$ with
\be
\label{eqn:xcmag}
h^2 = \frac{3}{4d} \left[ d -1 + \sqrt{\left(1-{\scriptstyle\third} {d}\right)
\left(1+d\right)} \ \right],
\ee
where we have defined $d \equiv \left( 1 + e^{-2 \beta}
\right)^p$. This gives $x_c$ in the magnetized phase---we shall prove
that it is magnetized later.

\subsection{Number of blobs and $\gst$}
\label{sec:blobgam}
Next we will calculate $\nmean_r$, the average number of 1PI blobs at
distance $r$. Define $G_r =f^{(r)}(v T)$, where the notation means
$f(f(f(...f(v T))))$;  this weights each blob at distance $r$ with an
extra factor of $v$. Note that $v T$ can be regarded as the standard
partition function, but with the root blob weighted by an extra factor
of $v$. Applying the function $f$ to this gives $f(v T)$, where now
each blob at distance one is weighted with the extra factor of
$v$. Each application of $f$ just moves the weights of $v$ 
down the tree, by one unit of distance. Thus,
\be
\nmean_r = \left[ \frac{v}{G_r} \frac{\partial G_r}{\partial v}
\right]_{v=1} = \frac{1}{T} \left.
\frac{\partial G_r}{\partial v} \right\vert_{v=1} ,
\ee
where we have used that $G_r(v \! = \! 1)=T$. However, $G_r=f(G_{r-1})$ and
thus
\begin{eqnarray}
\nmean_r &=& \frac{1}{T} \left[ \left. \frac{\partial f}{\partial y}
\right\vert_{y=G_{r-1}} \frac{\partial G_{r-1}}{\partial v} \right]_{v=1} \\
    &=& \left.\frac{\partial f}{\partial y} \right\vert_{y=T} \nmean_{r-1} 
\label{eqn:expo}
    = \left[ \left.\frac{\partial f}{\partial y}\right\vert_{y=T}
\right]^r ,
\end{eqnarray}
since $\nmean_0=1$ (there is only one root blob).
It will be convenient to define,
\be
B(x, \beta, p) = \left. \frac{\partial f}{\partial y} \right\vert_{y=T}.
\ee
Using the formula (\ref{eqn:f}) for $f(y)$, one can calculate $B$,
\be
\label{eqn:b}
B=\left. \frac{\partial f}{\partial y} \right\vert_T =
\frac{\lambda}{2}
\frac{\ztwo(x')}{ \sqrt{1-\lambda T}} \left( 3 \mtwo(x') -1 \right) + 2 x
\left( 1+e^{-2 \beta} \right)^{2p} T,
\ee
where
\be
\label{eqn:model2m}
\mtwo(x) \equiv \frac{x}{\ztwo(x)}\frac{\partial \ztwo}{\partial x} =
\frac{1-\tau}{1- 3 \tau}.
\ee
Note that $\mtwo(\xctwo)=\frac{5}{3}$. Evaluating at $x_c$ will
give $B_c(\beta,p)$. In the magnetized phase this gives us,
\be
\label{eqn:bc}
B_c= 1 + \half \left( d -1 \right)
- \frac{1}{2 \sqrt{3}} \sqrt{4 - \left( d -1 \right)^2} .
\ee
Thus at $x_c$, $\nmean_r=(B_c)^r$ and the average total number
of blobs is
\be
\nmean = \frac{1}{1 - B_c}.
\ee
For large enough $\beta$, $B_c$ is less than one and positive, that is, the
number of blobs decreases exponentially with distance. As $\beta$ is reduced,
$B_c$ increases (since $\lambda$ increases and this encourages the
tree to branch), until $B_c=1$ at some critical value, $\beta^*$; this
corresponds to the boundary of the tree-like region. Using $T=f(T)$,
one can show that,
\be
\frac{\partial T}{\partial x} = \frac{1}{1-B} \left[ 
d^2 T^2 + \frac{\ztwo(x')}{x \sqrt{1 - \lambda T}}
\left( \mtwo(x') +x d T \left( \mtwo(x') -1 \right) \right)
 \right].
\ee
For $\beta > \beta^*$, $0<B_c<1$ and $\left. \frac{\partial T}{\partial x}
\right\vert_{x_c}$ is
finite, whilst for $\beta \le \beta^*$, $\frac{\partial T}{\partial
x}$  diverges as $x \to x_c$. That is, $B_c=1$ throughout the
entire tree-like region and $\gst>0$ in this region, since
$\frac{\partial T}{\partial x} \sim (x_c -x)^{-\gst}$.
One can easily calculate $\gst$ in the various regions of the phase
diagram. In the tree-like region and on the boundary,
$\frac{\partial T}{\partial x} \sim  \frac{1}{1-B}$, hence $1-B \sim
{(\Delta x)}^{\gst}$, where $\Delta x \equiv x_c -x$. However, we also have an
expression (\ref{eqn:b}) for $B$ and this  gives $1-B \sim {(\Delta
x)}^{1-\gst}$ inside the tree phase. Hence, $\gst=\half$ here as expected.
On the boundary $1-B \sim {(\Delta x)}^{\half (1-\gst)}$ giving $\gst^*
=\third$. In the magnetized phase $\frac{\partial^2 T}{\partial x^2}
\sim \frac{\partial^2 \ztwo(x')}{\partial {x'}^2} \sim (\Delta
x')^{-\half} \sim (\Delta x)^{-\half}$, giving $\gst=-\half$ as expected.

Equation (\ref{eqn:bc}) applies in the magnetized phase and also on the
boundary between the two phases. At $\beta^*$, $B_c=1$ and this gives
using (\ref{eqn:bc}),
\be
\beta^*= - \half \ln \left( 2^{\frac{1}{p}} -1  \right).
\ee
Note that this gives an estimate for the location, in the full model,
of the transition between the tree and magnetized phases; compare this
with the estimated location of the transition from the
tree-like to the unmagnetized
(pure gravity) phase, given in~\cite{paper3}.

\subsection{$x_c$ in the tree phase}
Next we will calculate $x_c$ in the tree phase. At $x_c$ we have from
$T_c=f(T_c)$,
\be
\label{eqn:a1}
T_c = \sqrt{1 - \lambda_c T_c} \ \ztwo(x') + x_c \left(1+ e^{-2 \beta}
\right)^{2p} T_c^2
\ee
and from $B_c=1$, using (\ref{eqn:b}),
\be
\label{eqn:b1}
\frac{\lambda_c}{2} \frac{\ztwo(x')}{ \sqrt{1-\lambda_c T_c}}
\left( 3 \mtwo(x') -1 \right) + 2 x_c
\left( 1+e^{-2 \beta} \right)^{2p} T_c = 1 ,
\ee
where $x'=x_c (1- \lambda_c T_c)^{-\frac{3}{2}}$. Remembering that
$d=\left( 1+ e^{-2 \beta} \right)^p$ and defining $X \equiv 1 -
\lambda_c T_c$, (\ref{eqn:b1}) gives
\be
\ztwo \left(3\mtwo -1\right) x' X d + 2 d^2 T_c x' X^{\frac{3}{2}} =1
\ee
and (\ref{eqn:a1}) gives
\be
\label{eqn:c1}
d^2 T_c^2 x' X^{\frac{3}{2}} = T_c - X^{\half} \ztwo .
\ee
Thus,
\be
\ztwo \left( 3\mtwo -1 \right) x' X d +1 - \frac{2}{T_c} X^\half \ztwo
=0 .
\ee
Moreover, 
\be
T_c = \frac{1-X}{\lambda_c} = \frac{1-X}{2 x' X^\frac{3}{2} d}
\ee
and thus 
\be
\ztwo \left(3 \mtwo -1 \right) X(1-X) x' d + 1-X - 4 x' \ztwo X^2 d
=0.
\ee
Also (\ref{eqn:c1}) gives
\be
\label{eqn:d1}
4x' \ztwo X^2 d = (1-X) \left[ 2 - d(1-X) \right].
\ee
Combining these last two equations gives
\be
 \left[ \ztwo (3 \mtwo -1) x' -1 \right] X d + d-1=0.
\ee
Using (\ref{eqn:model2z}) and (\ref{eqn:model2m}),
\be
X=\frac{d-1}{d(1-2 \tau)}
\ee
and substituting into (\ref{eqn:d1}) gives
\be
\tau = \half \left[1+2(d-1)^2 \right]^{-1}
\ee
and thus $x_c=x' X^{\frac{3}{2}}$ is given by
\be
\label{eqn:xctree}
x_c = \half d^{-\frac{3}{2}} \sqrt{d-1} 
\ee
in the tree phase. From equations (\ref{eqn:xctree}) and
(\ref{eqn:xcmag}) one can easily show that there is a third
order phase transition with a finite discontinuity, that is, the
critical exponent $\alpha=-1$.

\subsection{Number of vertices}
One can also calculate the average number of vertices at distance
$r$. Define $G'_r =f^{(r)}\left( f(T)\vert_{xz} \right)$, where the
notation $f(y)\vert_{xz}$ means the function $f(y)$, but with $x$
replaced everywhere by $xz$. Thus $f(T) \vert_{xz}$ is just a tree,
but with each vertex in the root blob weighted by an extra factor of
$z$. Applying $f$, $r$ times, just pushes these weights down the tree to
a distance $r$, so that in $G'_r$ all the vertices at distance $r$
have an extra weight of $z$. Thus the average number of vertices at
distance $r$ is given by
\begin{eqnarray}
\Nmean_r &=& \left[ \frac{z}{G'_r} \frac{\partial G'_r}{\partial z}
\right]_{z=1} = \frac{1}{T}\left[ \left. \frac{\partial f}{\partial y}
\right\vert_{y=G'_{r-1}} 
\frac{\partial G'_{r-1}}{\partial z}
\right]_{z=1} \\
\label{eqn:nmeanr}
&=& B^r \Nmean_0 
\end{eqnarray}
and this is essentially just the (geometric) two-point function (see
section~\ref{sec:dh}).
The average total number of vertices is 
\be
\Nmean = \frac{ \Nmean_0}{1-B} \ ,
\ee
where $\Nmean_0$ is the average number of vertices in the root
blob. This is given by
\be
\Nmean_0 = B + \mtwo(x') - x d^2 T \left( \mtwo(x') +1 \right).
\ee
In the magnetized phase this gives, at $x_c$,
\be
\Nmean_0 = \frac{4}{3} + \frac{1}{6} (d-1) + \frac{1}{2 \sqrt{3}}
\sqrt{4 - (d-1)^2}
\ee
and in the tree phase,
\be
\Nmean_0 = \frac{2 (d-1)}{(2d-3)} .
\ee
Note that $d \equiv (1+e^{-2 \beta})^p$, with $1<d<2$ in the magnetized
phase and $d > 2$ in the tree phase. The formulae also show that the
average number of vertices in a blob is very small, for any values of
$\beta$ and $p$; in fact $\Nmean_0=2$ at $\beta^*$.

\section{Critical exponents from the scaling relations}
\label{sec:toycrit}
{}From equation (\ref{eqn:expo}) we see that there is an exponential
decay of the number of blobs with distance.
One can define a mass, $m$, in the model through,
\be
\label{eqn:a2}
\nmean_r = B^r = e^{-m r}.
\ee
The mass is a function of $\beta$, $p$ and $x$.
At $x_c$, as the critical line is approached from the magnetized
phase, the mass vanishes (i.e.\ there is a correlation length equal to
$1/m$, which diverges). Note that at $x_c$, $m=0$ throughout the whole
tree phase.
Let us use this definition of the mass and various scaling
relations to calculate the critical exponents. Later we will calculate
the magnetic exponents directly from the model, for the case $p=1$,
gaining the same results for $\beta_m$ and $\delta_m$; this
will show that our definition of the mass is reasonable and that the
scaling relations hold for this toy model.

First let us consider the geometric exponents; note that we have already
calculated $\gst$ in section~\ref{sec:blobgam}. 
The exponent $\nu$ is defined by $m \sim
(\Delta x)^\nu$ for $\Delta x \to 0$ (where $\Delta x \equiv x_c-x$),
but $m=-\ln B$ so that $ m \sim 1- B \sim
{(\Delta x)}^{\gst}$ and hence $\nu=\gst$; in the tree phase
$\nu=\gst=\half$ and on the boundary $\nu^*=\gst^*=\third$.
Since there are no power law corrections to (\ref{eqn:a2}) we have
$\eta=1$; note that at small $\Delta x$, for $1 \ll r \ll 1/m$, 
one might expect~\cite{AmbWat95} that
$\nmean_r \sim r^{1-\eta}$. These sets of exponents satisfy
Fisher's scaling relation $\gst = \nu (2- \eta)$.

In reference~\cite{AmbWat95} it is shown that if the two-point function has
associated with it a vanishing mass, then
for a suitable definition of the Hausdorff dimension, $\nu d_H
=1$. This gives that $d_H=2$ in the tree phase, as we might expect for
branched polymers and $d_H^*=3$ on the critical line.

Consider now the magnetic exponents, which we will write with a
subscript $m$ to avoid confusion. Evaluating at $x_c$ and letting
$\beta \to \beta^*$ from the magnetized phase, we have, defining
$\Delta \beta \equiv \beta - \beta^*$, 
$m \sim \Delta \beta$ from (\ref{eqn:bc}) and hence $\nu_m=1$. If
we take $\eta_m=1$ (we shall see later that the spin-spin correlation
function has no power law corrections either), then using
$\gamma_m = \nu_m (2- \eta_m)$ gives $\gamma_m=1$. Applying the other
scaling relations $2- \alpha= \nu_m d_H^*$, $\beta_m \delta_m =
\beta_m + \gamma_m$ and $\alpha +2 \beta_m + \gamma_m =2$, yields
$\alpha = -1$, $\beta_m=1$ and $\delta_m=2$.
The various exponents are listed in table~\ref{tab:exp}.

\begin{table}[htb]
\caption{Critical exponents}
\label{tab:exp}
\begin{center}
\begin{tabular}{|c||c|c|c|c|} \hline
Phase      & $\gst$   & $\nu$    & $\eta$ & $d_H$ \\ \hline
Magnetized & $-\half$ &          & (1)      &   \\ \hline
Critical   & $\third$ & $\third$ & 1      & 3 \\ \hline
Tree       & $\half$  & $\half$  & 1      & 2 \\ \hline
\end{tabular}
\begin{tabular}{c}
\hphantom{xxxx}\\
\end{tabular}
\begin{tabular}{|c|c|c|c|c|c|}\hline
$\alpha$ & $\beta_m$ & $\gamma_m$ & $\delta_m$ & $\nu_m$ & $\eta_m$ \\ \hline
$-1$ & 1 & 1 & 2 & 1 & 1 \\ \hline
\end{tabular}
\end{center}
\end{table}

Before proceeding to check the magnetic exponents explicitly, we shall
define $d_H$ and calculate it directly from the model.

\section{Hausdorff dimension}
\label{sec:dh}

As before in~\cite{AmbWat95} we define the Hausdorff dimension, $d_H$, 
in terms of the two-point
function
\be
T_2(r) = \sum_{ G \in \gmark} x^N W_G,
\ee
where $\gmark$ is the set of planar \ph3 graphs,
with two marked points that are
separated by a distance~$r$. One of the marked points will be taken
to be the vertex, in the root blob, which is connected to the external
leg. $W_G$ is just the usual weight for the spin configurations, which
appears in~(\ref{eqn:gcpf}). Then $d_H$ is defined, in the continuum
limit, by
\be
N(r) \sim r^{d_H}, \mgap r \to \infty, \mgap m(\Delta x) r = {\rm
const.},
\ee
where
\be
N(r) \equiv \frac{1}{T_2(r)} \sum_{G \in \gmark} N \ x^N \ W_G = x
\frac{\partial \left( \ln T_2(r) \right)}{\partial x} .
\ee
That is, by tuning $x$ to
$x_c$ we are taking a continuum limit; however 
this definition of $d_H$ only really makes sense in the tree phase and on the
boundary, where the mass vanishes at $x_c$.

Now $T_2(r)$ is just
\be
T_2(r) = \sum_{G \in \gone} x^N W_G N_r = T \Nmean_r = T B^r \Nmean_0,
\ee
where $N_r$ is the number of vertices at distance $r$ for graph $G$;
\be
N(r) = x \left[ \frac{1}{T} \frac{\partial T}{\partial x} +
\frac{r}{B} \frac{\partial B}{\partial x} +
\frac{1}{\Nmean_0} \frac{\partial \Nmean_0}{\partial x} \right] .
\ee
As $r \to \infty$,
\be
N(r) \sim r  \frac{\partial \mtwo}{\partial x'}  
\frac{\partial T}{\partial x} \sim r \left( \xctwo - x'
\right)^{-\half} \left(\Delta x \right)^{-\gst}.
\ee
Since we are taking the continuum limit with $m r$ fixed, $r
\sim m^{-1} \sim (\Delta x)^{-\gst}$,
\be
N(r) \sim r^2 \left( \xctwo - x' \right)^{-\half}.
\ee
In the tree phase $N(r) \sim r^2$ giving $d_H=2$ and on the
critical line $N(r) \sim r^2 (\Delta x')^{-\half} \sim r^2
(\Delta x)^{- \third} \sim r^3$, so that $d_H^* =3$, as
expected. Unfortunately, due to the way distances and the Hausdorff
dimension have been defined in this model, $d_H$ is not well-defined in
the magnetized phase. In reference~\cite{AmbWat95} it is shown, using
the geodesic distance, that $d_H=4$ for
pure gravity (also the geometric exponents are $\nu=1/4$ and $\eta=4$)
and this may well be true in the whole of the magnetized
phase. It is certainly the correct value at $p=0$, where the toy model
reduces to a pure gravity model.

Thus our definition of distance appears to correctly capture the
behaviour of the model in the tree phase and on the critical line,
where $\gst>0$, but not within the magnetized phase.

\section{Spin-spin correlation function}
\label{sec:spinspin}
In this section we will calculate $\smean_r$ the average total spin at
distance $r$; the summation is over all vertices at distance $r$ and
all spin flavours on those vertices. Since the spins in the root blob
are fixed to be $+1$, this is essentially a spin-spin correlation
function, for spins separated by a distance $r$.
To calculate this we will add a magnetic field for vertices at
distance $r$. First we will solve the model with $p=1$ and then the
general case. Let us define
\be
f_+(y) = \sqrt{1- \lambda y e^H} \ \ztwo \left( x e^H \left( 1 - \lambda
y e^H \right)^{-\frac{3}{2}} \right) + x (1+e^{-2 \beta})^2 y^2 e^H,
\ee
with $\lambda = 2 x (1+e^{-2 \beta})$. This is just $f(y)$, but with
$x$ replaced by $x e^H$. The function takes a number of trees with
weights $y$ and glues them on to a blob, where each vertex of the blob
has a spin of $+1$ on it and there is a magnetic field $+H$ applied on
these spins. The function $f_-(y)$ is defined in the same way, but
with $H$ replaced by $-H$ throughout. 

Let $T^+_r$ be a tree with positive spins in the root blob and a
magnetic field of $+H$ applied at distance $r$ from the root. Similarly
$T^-_r$ has spins of $-1$ in the root blob. Then $T^+_0 =f_+(T)$ and
$T^-_0=f_-(T)$. Define,
\be
F(y)= \sqrt{1-2xy} \ \ztwo \left( x \left(1-2xy\right)^{-\frac{3}{2}}
\right) + x y^2 .
\ee
Then
\begin{eqnarray}
T^+_r &=& F(T^+_{r-1} + e^{-2 \beta} T^-_{r-1}) \\
T^-_r &=& F(T^-_{r-1} + e^{-2 \beta} T^+_{r-1})  ,
\end{eqnarray}
that is, to make a graph with spins of $+1$ on the root blob and a
magnetic field at distance $r$, one takes a number of trees with
magnetic fields at distance $r-1$ and glues them on to a blob with
positive spins, picking up a factor of $e^{-2 \beta}$ if the spins on
the root of the tree are negative. Now,
\be
\smean_r^+ = \left[ \frac{1}{T^+_r} \frac{\partial T_r^+}{\partial H}
\right]_{H=0}  = \left. \frac{1}{T} \frac{\partial T_r^+}{\partial H} 
\right\vert_{H=0} ;
\ee
the superscript on $\smean_r^+$ denotes the fact that the vertices in
the root blob have a single spin each, which is fixed to be $+1$.
\be
\frac{\partial T_r^+}{\partial H} =
\left[ \frac{\partial T_{r-1}^+}{\partial H} + e^{-2 \beta} 
\frac{\partial T_{r-1}^-}{\partial H} \right]
\left. \frac{\partial F}{\partial y} \right\vert_{y=T_{r-1}^+ + e^{-2
\beta} T_{r-1}^-}
\ee
\be
\smean_r^+ =
\left[ \smean_{r-1}^+ + e^{-2 \beta} \smean_{r-1}^- \right]
\left. \frac{\partial F}{\partial y} \right\vert_{y=\left(1 + e^{-2
\beta}\right) T} 
\ee
but $\smean_{r-1}^- = - \smean_{r-1}^+$ and $F(y)=f(2xy/\lambda)$ so
that
\be
\left. \frac{\partial F}{\partial y} \right\vert_{y=\left(1 + e^{-2
\beta}\right) T} = \frac{2xB}{\lambda}.
\ee
Thus defining $t\equiv \tanh \beta$,
\be
\smean_r^+ = B t \ \smean_{r-1}^+ = (B t)^r \smean_0^+ =(Bt)^r \Nmean_0.
\ee
Let us consider the general case of $p$ spin flavours for which
$\lambda = 2 x \left(1 + e^{-2 \beta} \right)^p$,
$$
T_r^{++ \cdots +}=F\Bigl(T_{r-1}^{+\cdots+} + e^{-2 \beta} \left(
T_{r-1}^{+\cdots+-} + T_{r-1}^{+\cdots+-+} + \cdots \right) +
$$
\be
 \mgap
e^{-4 \beta} \left( T_{r-1}^{+\cdots+--} + \cdots \right) + \cdots + e^{-2
\beta p} T_{r-1}^{-\cdots-} \Bigr)
\ee
\be
\smean_r^{++\cdots+} = \frac{2xB}{\lambda} \left[
\smean_{r-1}^{+\cdots+} + e^{-2 \beta} \left(
\smean_{r-1}^{+\cdots+-}  + \cdots \right)
+ \cdots + e^{-2 \beta p} \smean_{r-1}^{-\cdots-} \right] .
\ee
However one can easily show that,
\be
\smean_r^{\cdots} = \frac{1}{p}(p-2n) \smean_r^{++\cdots+},
\ee
where there are $p \! - \! n$ plus signs and $n$ minus signs in the superscript
on the left-hand side. Thus,
\begin{eqnarray}
\smean_r^{++\cdots+} &=& \frac{B}{\left(1+e^{-2 \beta} \right)^p}
\smean_{r-1}^{+\cdots+} \sum_{n=0}^p e^{-2 \beta n} \frac{1}{p} (p-2n)
\left( p \atop n \right) \\
&=& (Bt) \smean_{r-1}^{++\cdots+} = (Bt)^r p \Nmean_0 .
\end{eqnarray}
Now defining the magnetization at distance $r$ by
\be
{\cal M}_r \equiv \frac{ \smean_r^{+ \cdots +}}{p \Nmean_r} ,
\ee
we have that ${\cal M}_r = t^r$. 
It should be noted that if one uses the exponential decay of
$\smean_r^{+ \cdots +}$ or ${\cal M}_r$ to define a spin-spin
correlation length, then this quantity will not diverge at the phase
transition; this behaviour should be contrasted with that of the
geometric correlation length, which does diverge.

If we define the total magnetization
${\cal M}$, for the case in which all the spins in the root blob are
positive, by
\be
\label{eqn:mgce}
{\cal M} \equiv \frac{\smean^{+\cdots+}}{p \Nmean},
\ee
where $\smean^{+\cdots+}$ is the average of the total spin, which is a sum over
all vertices and all spin flavours, then we have, at $x_c$,
\be
{\cal M} = \frac{1-B_c}{1-B_ct} .
\ee
Note that ${\cal M}=1$ for $\beta = \infty$, and that ${\cal M}=0$
throughout the tree phase, showing that this phase is
unmagnetized. The other phase has $0<B_c<1$ and hence $0<{\cal M}\le 1$;
this is the magnetized phase as claimed earlier. Also at $x_c$, near
$\beta^*$, $1-B_c \sim \Delta \beta$, where $\Delta \beta \equiv \beta
- \beta^*$, so that ${\cal M} \sim \Delta
\beta$ and thus $\beta_m=1$, as calculated from the scaling relations.
In fact for finite $p$,
\be
{\cal M} = \frac{4p}{3} \Delta \beta + O((\Delta \beta)^2),
\ee
so that the coefficient in front of $\Delta \beta$ is non-zero in
general (note that $p=0$ would give $\beta^*=-\infty$).

\section{Toy model, with a magnetic field}
\label{sec:toymag}
In the previous section we defined the magnetization in the grand
canonical ensemble by (\ref{eqn:mgce}).
A different definition is possible namely,
\be
{\cal M}_{ce} = \lim_{N \to \infty} \frac{1}{pN} \smean_N,
\ee
where now we are working in the canonical ensemble, that is, we are
using the set, $\gone(N)$ of rooted $N$-vertex graphs.

In this section we will add a magnetic field, $H$, to the toy
model. However, we shall only consider the case $p=1$, as $p>1$ appears
to be more difficult to solve. It will be shown that the magnetic
exponents $\beta_m$ and $\delta_m$ are the same for both definitions
of the magnetization and agree with the results from the scaling
relations.

Rather than use the transfer function $f(y)$, which glues trees on to a
1PI blob, we shall use a different transfer function $\fbar(y)$,
that glues trees on to a domain; this is more convenient for the $p=1$
model with a magnetic field. For $H=0$, it will give a sum over 
the same set of graphs that we had previously, with the same weights, 
but we will no longer be
able to keep track of distances within the graphs.

Consider first the case $H=0$, the function $\fbar(y)$ is given by
\be
\label{eqn:fbar}
\fbar(y)= \frac{1}{\sqrt{1-2xK}} \ \zone \left( x \left(1-2x K
\right)^{-\frac{3}{2}} \right) + x K^2 ,
\ee
where
\be
K= \frac{1}{2x} \left( 1 - \sqrt{1- \mu y} \right) ,
\ee
with $\mu = 4 x e^{-2\beta}$ and
\be
\zone(x)=\sum_{G \in \gone} x^N .
\ee

The function defined in (\ref{eqn:fbar}) takes a rooted \ph3 graph,
which will form the core of a domain, and glues on trees weighted with
factors of $K$. As before one picks up a factor of $2x$ for each tree
glued on. In this case however we allow trees to be glued on to the
root link, which changes the power of the factor in front of $\zone$
compared with that in (\ref{eqn:f}). This is necessary to make sure
that we correctly sum over all possible domain structures.

The factor $K$ is the solution of
\be
K= e^{-2\beta} y + x K^2 .
\ee
This generates tree graphs whose vertices are weighted with $x$ and
which have a factor of $e^{-2\beta} y$ at the end of each branch. The
domain corresponds to the core \ph3 graph plus all the vertices in
$K$. The ends of the branches in $K$, which we have weighted with
$e^{-2\beta} y$, correspond to the domain boundaries; setting $y=T$
will give us a set of domains glued together in a tree-like fashion.

\begin{figure}[bth]
\caption[l]{$T=\fbar(T)$}
\label{fig:domain}
\begin{picture}(100,55)(0,0)
\centerline{\epsfbox{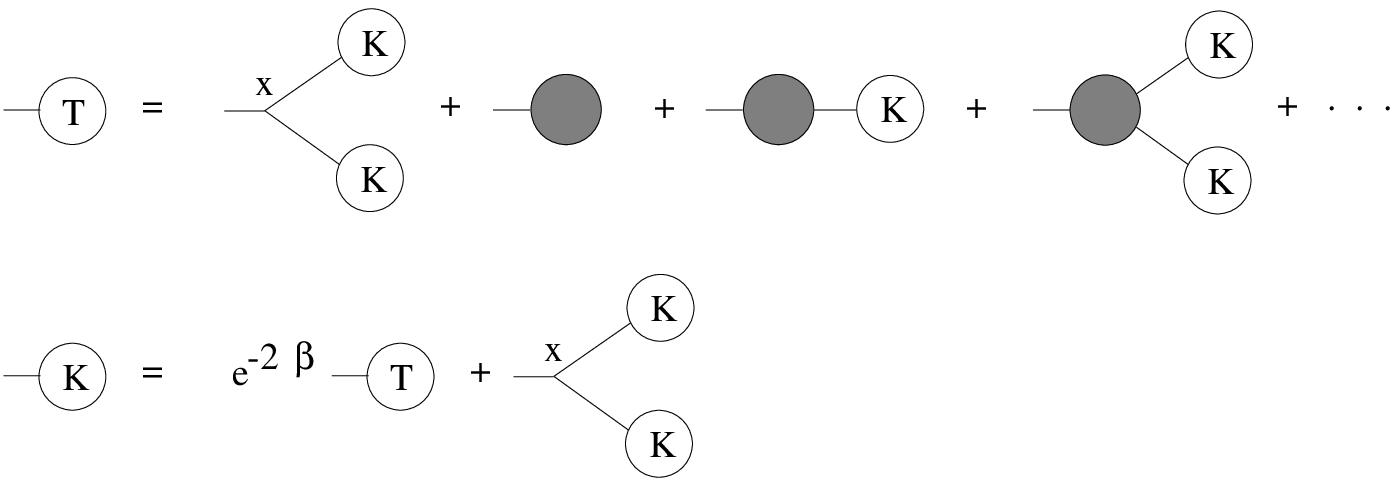}}
\end{picture}
\end{figure}

Thus we have
\be
\fbar(y) = \frac{1}{(1- \mu y)^{\quart}} \ \zone \left( x \left( 1 -
\mu y \right)^{- \frac{3}{4}} \right) + \frac{1}{4x} \left( 1
- \sqrt{1- \mu y} \right)^2
\ee
and the grand canonical partition function (\ref{eqn:gcpf}) is the
solution of $T= \fbar(T)$. This is shown diagrammatically in
fig.~\ref{fig:domain}.

Note that this is just a different way of formulating the same model
that we solved previously. One can easily use $T=\fbar(T)$ to
calculate $x_c$ in the two phases, gaining the same results as before
(to solve the $p$-flavour case one just uses $\mu=4x(d-1)$). The
advantage of this new formulation is that as we use $\fbar$ to move
down the tree from one domain to the next, the sign of the spins
alternates (at least for $p=1$).
This simplifies greatly the task of adding a magnetic field.
For the original version of the model it is quite
difficult to keep track of the spins on the different blobs.

Now let us add a magnetic field of $+H$, the transfer function which
makes a domain with spins of $+1$ is
\be
\label{eqn:fbarplus}
\fbar_+(y) = \frac{1}{(1- \mu y e^H)^{\quart}} \ \zone \left( x e^H \left( 1 -
\mu y e^H \right)^{- \frac{3}{4}} \right) + \frac{1}{4x e^H} \left( 1
- \sqrt{1- \mu y e^H} \right)^2 .
\ee
Each spin in the domain is given an extra weight of $e^H$.
It will be convenient to define
 $x_+'=x e^H  \left( 1 - \mu y e^H \right)^{- \frac{3}{4}}$.
Similarly $\fbar_-(y)$, which makes a domain with negative spins is 
(\ref{eqn:fbarplus}), but with $H$ replaced everywhere by $-H$.
Let $T^+$ be the partition function for graphs with positive spins in
the root domain and $T^-$ that for those with a negative root domain.
Then $T^+=\fbar_+(T^-)$ and $T^-=\fbar_-(T^+)$. These equations
determine $T^+(x,\beta,H)$. For $H>0$ in the magnetized phase, the
critical value of $x$ is determined by $x_+' = \xcone$, where $\xcone$ is the
critical value for $\zone(x_+')$. Defining
\be
\hbar \equiv \left( \frac{x_c e^H}{\xcone} \right)^\third,
\ee
we have
\be
T^-_c = \frac{1}{\mu_c e^H} \left( 1 - \hbar^4 \right)
\ee
and
\be
T^+_c = {\hbar}^{-1} \zone_c + \frac{1}{4 x_c e^H} \left( 1 - \hbar^2
\right)^2 .
\ee
Note that $\xcone=1/(2.3^{\frac{3}{4}})$, $\zone_c=3^\frac{3}{4}
\left( 1 - \frac{\sqrt{3}}{2} \right)$ and $\mone_c =
1/(\sqrt{3} -\frac{3}{2}) -1 $ from~\cite{BIPZ}. However we still
have $T^-=\fbar_-(T^+)$, and using this gives an equation determining
$\hbar$. Putting
\be
X \equiv 1 - \mu_c e^{-H} T^+_c = 1 - e^{-2\beta} e^{-2H} \left( 1 -
\sqrt{3} \hbar^2 + \hbar^4 \right)
\ee
and $x'=x_c e^{-H} X^{-\frac{3}{4}}$, then we have
\be
\label{eqn:hbar}
e^{2 \beta} (1- \hbar^4) = e^{2H} \left[ 1 + X + 2 \sqrt{X} \left( 2
x' \zone(x')-1 \right) \right].
\ee
This gives $\hbar(\beta,H)$ and hence $x_c(\beta,H)$.
Now,
\be
\mce(\beta,H) = - \frac{1}{x_c} \frac{\partial x_c}{\partial H} = 1 -
\frac{3}{\hbar} \frac{\partial \hbar}{\partial H} .
\ee
At $H \to 0$ we get by differentiating (\ref{eqn:hbar}),
\be
\mce(\beta,H \to 0^+) = 1 - \frac{3 \left(2 - \frac{2}{\sqrt{3}} \hbar^2 -
\sqrt{3} (\hbar)^{-2} \right)}{\left( 3 - \hbar^2 \sqrt{3} -
\hbar^4 \left(2 e^{2 \beta} + e^{-2\beta} +3 \right) \right)} ,
\ee
with
\be
\hbar^2 = \frac{1}{2} \frac{1}{\left(1+e^{-2\beta}\right)} \left[
\sqrt{3} e^{-2 \beta} + \sqrt{4 - e^{-4 \beta}} \right] ;
\ee
note that at $H=0$, $\hbar^2 = 2 h^2 / \sqrt{3}$, see equation
(\ref{eqn:xcmag}). At $\beta$ close to $\beta^*=0$,
\be
\mce(\Delta \beta , H \to 0^+ ) = 2 \Delta \beta + O((\Delta \beta)^2).
\ee
Hence, $\beta_m =1$ as previously, but note that $\mce$ is not equal
to ${\cal M}$. Now consider $\delta_m$, defined by $\mce(\beta^*,
\Delta H) \sim (\Delta H)^{\frac{1}{\delta_m}}$. From (\ref{eqn:hbar}),
one can show that $\mce$ is a certain function of $\hbar$, $e^{-2
\beta}$, $e^{2 H}$, $x'$, $\zone(x')$, $\mone(x')$ and $X$, where
\be
\mone(x') \equiv \frac{x'}{\zone(x')}\frac{\partial \zone(x')}{\partial x'}.
\ee
At $\beta^*$ and small $\Delta H$, barring accidental cancellations,
then all these functions are a constant plus terms of order $\Delta H$
(or smaller), except for $\mone(x')$,
\be
\mone(x') = \mone_c + \left(\xcone - x' \right)^\half + \cdots .
\ee
Since $x' = \xcone + O(\Delta H)$ then $\mone(x') = \mone_c + O((\Delta
H)^\half)$, giving $\delta_m =2$. Note that the value of $\delta_m$
depends crucially on the fact that $\gst=-\half$ in the magnetized
phase.

Suppose that we next try to calculate $\gamma_m$, defined by
\be
\chi(\beta,H=0) \sim (\Delta \beta)^{-\gamma_m},
\ee
then we find that $\chi \sim \frac{\partial^2 x_c}{\partial H^2}$ and
that this contains terms such as $\frac{\partial^2 \zone(x')}{\partial
{x'}^2}$, which diverge as we take $H$ to zero. Thus $\gamma_m$ seems
not to be well-defined, even though the scaling relations give
$\gamma_m=1$.

Consider now the magnetization in the grand canonical ensemble,
${\cal M}$, defined by (\ref{eqn:mgce}), $\beta_m$ has
already been calculated and this just leaves $\delta_m$. We see that
${\cal M} \to 0$ at $\beta^*$, due to the divergence of $\Nmean$,
\be
{\cal M}(\beta^*,\Delta H) \sim \left. \frac{1}{\Nmean}
\right\vert_{\beta^*} .
\ee
For graphs with a positive root domain, at $x_c$,
\be
\Nmean^+ = \frac{x_c}{T^+_c} \left. \frac{\partial T^+}{\partial x}
\right\vert_{x_c} ,
\ee
where $T^+=\fbar_+(\fbar_-(T^+))$.
This gives 
\be
\frac{\partial T^+}{\partial x} \left[ 1-
\left. \frac{\partial \fbar_+}{\partial y} \right\vert_{T^-}
\left. \frac{\partial \fbar_-}{\partial y} \right\vert_{T^+} \right] =
\left. \frac{\partial \fbar_+}{\partial x} \right\vert_{T^-} +
\left. \frac{\partial \fbar_+}{\partial y} \right\vert_{T^-}
\left. \frac{\partial \fbar_-}{\partial x} \right\vert_{T^+} .
\ee
Thus 
\be
{\cal M}(\beta^*, \Delta H) \sim  1-
\left. \frac{\partial \fbar_+}{\partial y} \right\vert_{T^-_c}
\left. \frac{\partial \fbar_-}{\partial y} \right\vert_{T^+_c}
\ee
and again this is a function of various variables all of which are
a constant plus $O(\Delta H)$, except that $\mone(x')=\mone_c + O((\Delta
H)^\half)$, giving $\delta_m=2$.

This completes our analysis of the toy model; we have derived all the
critical exponents from the model (except for $\gamma_m$) and shown
that all the usual scaling relations hold. Now we shall turn our
attention to the full model of $p$-Ising spins coupled to 2d gravity,
and will find that many of the considerations in the previous sections
apply for the transition between the tree and magnetized phases in
this model.

\section{Full model}
\label{sec:full}
\subsection{Definition}
Consider the full model where we have $p$ independent Ising spins on each
vertex and are summing over all rooted planar
\ph3 graphs and all spin configurations.
Then the partition function is, with $t=\tanh \beta$,
\be
\label{eqn:full}
\zone(x,\beta) = \sum_{G \in \gone} x^N \left[ \frac{1}{2^N}
\sum_{\{S\}} \prod_{<ij>} \left(1+t S_i S_j \right) \right]^p ,
\ee
where we have divided out various factors of two and $\cosh \beta$ in
order to simplify the formulae later on. $N$ is the number of vertices
in graph $G$. Note that there is no vertex and are no spins at the end
of the root, so that the product over links does not include the root
link. 
In fact we shall consider a generalization of this model that has $p$
coupling constants. Expanding out the above equation gives, with the
extra coupling constants,
\be
\label{eqn:general}
\zone(x,\{\lambda\}) = \sum_{G \in \gone} x^N \frac{1}{2^{Np}}
\sum_{\{S\}} \prod_{<ij>} \biggl[ 1 + \lambda_1 \sum_{\alpha=1}^{p}
S_i^\alpha S_j^\alpha + \lambda_2 \sum_{\alpha=1}^{p-1}
\sum_{\beta=\alpha+1}^{p} S_i^\alpha S_j^\alpha S_i^\beta S_j^\beta + \cdots
\ee
$$\mgap 
+ \lambda_p \prod_{\alpha=1}^{p} S_i^\alpha S_j^\alpha \biggr],$$
where $S_i^\alpha$ is the flavour $\alpha$ spin on vertex $i$, the
second summation is over all spin configurations and
$\{\lambda\}$ is the set of coupling constants
$\{\lambda_1,\lambda_2,\cdots,\lambda_p\}$, with $ 0 \le \lambda_j \le
1$ for all~$j$. 
Of course by setting $\lambda_j =t^j$ one recovers the ordinary model
(\ref{eqn:full}). Alternatively by setting $\lambda_j=0$ for $j>1$ we
recover the O(n) models studied in~\cite{onmodels,charlotte} with $n=p$.
To save writing we define $T \equiv
\zone(x,\{\lambda\})$. We will show that $T$ satisfies $T=f(T)$ where
\be
\label{eqn:fullf}
f(y)=\sqrt{1-2xy} \ \ztwo(x',\{\lambda'\}) + x y^2 ,
\ee
with 
\be
\label{eqn:fullmap}
x'=x(1-2xy)^{-\frac{3}{2}} , \mgap
\lambda_j' =\lambda_j \frac{(1-2xy)}{(1-2xy \lambda_j)} ;
\ee
$\ztwo(x',\{\lambda'\})$ is defined as in (\ref{eqn:general}),
but using the set of rooted 1PI graphs, $\gtwo$. In a similar fashion
to (\ref{eqn:f}) the function $f(y)$ takes a 1PI graph and glues trees
on to it, but this time as well as $x$ being changed to $x'$, the
coupling constants are also renormalized. Note that a similar
renormalization is studied in~\cite{Dur94}.

To justify the above equations let us first consider the case $p=1$.
For each link we have a factor such as $(1+\lambda_1 S_i S_j)$ and
when we multiply these factors together, the sum over spins will cause
any terms containing odd powers of a given spin to vanish. The only
non-vanishing terms correspond to sets of closed loops; we shall refer
to such a non-vanishing term as a loop configuration. Thus for each
graph $G$, we sum over all ways of drawing sets of closed
non-intersecting, non-backtracking loops on the graph (call this set
of loop configurations $\setloop$). So,
\be
\zone(x,\lambda_1) = \sum_{G \in \gone} x^N \sum_{\lcal \in \setloop}
\lambda_1^l ,
\ee
where $N$ is the number of vertices in graph $G$, and $l$ is the number
of links making up the loops in the loop configuration $\lcal$.

As before, we will make the graphs $G$ (where $G \in \gone$) by gluing
together 1PI blobs (fig.~\ref{fig:tree}). Note that because of the
tree-like structure, any given closed loop is wholly contained within
a single blob. This means that the partition function for a given
graph $G$, factorizes into a product of contributions from the
individual blobs. To calculate the transfer function, $f(y)$, we take a
rooted 1PI blob, summing over all graphs and loop configurations, as
well as over all ways of attaching trees to the blob. If no trees were
attached, one would just have $\ztwo(x,\lambda_1)$, which is
\be
\ztwo(x,\lambda_1) = \sum_{G \in \gtwo} x^N \frac{1}{2^N} \sum_{\{S\}}
\prod_{<ij>} \left[ 1 + \lambda_1 S_i S_j \right] =
\sum_{G \in \gtwo} x^N \sum_{\lcal \in \setloop} \lambda_1^l \ 1^{L-l},
\ee
where $L$ is the number of internal links in graph $G$ (i.e.\ excluding
the root link); that is, $L=\half (3N-1)$.
Attaching a tree weighted with $y$, gives a factor of $2xy$ (there is
an extra vertex and there are two ways of hanging the tree off the
link). For a given $G$ and $\lcal$, each link either contributes
$\lambda_1$ if it is part of a closed loop or $1$ if it is
not. Attaching an arbitrary number of trees to an internal
link that contributed
$1$ changes this contribution to $(1-2xy)^{-1}$; however for those
that contributed $\lambda_1$ we get $\lambda_1 (1-2xy
\lambda_1)^{-1}$, since adding a vertex on a link which was part of a
closed loop, increases the length of that loop by one and hence gives
an extra factor of $\lambda_1$. Thus summing over all ways of
attaching trees causes the change:
\be
(1+\lambda_1 S_i S_j) \longrightarrow \frac{1}{1-2xy} \left( 1 + S_i
S_j \frac{\lambda_1 (1-2xy)}{(1-2xy \lambda_1)} \right)
\ee
and hence
\be
\ztwo(x,\lambda_1) \longrightarrow \sqrt{1-2xy} \  \ztwo(x',\lambda_1').
\ee
The extra term in (\ref{eqn:fullf}) comes from the case in which the blob
is just a single vertex.

The general case for $p>1$ follows with only minor modifications to
the argument. Each closed loop is labelled with a spin flavour (an
integer $\alpha$, with $1\le \alpha \le p$) and
loops with different spin flavours can intersect. Thus for a given
graph $G$ and loop configuration $\lcal$, a given link will have $j$
spin flavours running through it ($0 \le j \le p$), giving a
corresponding factor of $\lambda_j$. Adding an arbitrary number of
trees to this link changes the contribution to $\lambda_j (1-2xy
\lambda_j)^{-1}$. Hence, we get (\ref{eqn:fullf}) and
(\ref{eqn:fullmap}) for the general case of a rooted 1PI blob with
trees, each weighted by $y$, hanging off it. Setting $y=T$, we recover the
partition function for graphs with $G \in \gone$, that is, $T=f(T)$;
$f(y)$ is the transfer function for the full model.

\subsection{Exponential decay of blobs}

As before, we define distance to be the geodesic distance, but count
only the links connecting 1PI blobs. Again one has $G_r =
f^{(r)}(vT)$, giving the partition function for a model where all blobs
at distance $r$ are weighted by $v$. The average number of 1PI blobs at
distance $r$ is
\be
\nmean_r = \left[ \frac{v}{G_r} \frac{\partial G_r}{\partial v}
\right]_{v=1} = \frac{1}{T} \left. \frac{\partial G_r}{\partial v}
\right\vert_{v=1} .
\ee
Since $G_r=f(G_{r-1})$,
\be
\nmean_r =\frac{1}{T} \left. \frac{\partial f}{\partial y}
\right\vert_{y=T} \left.\frac{\partial G_{r-1}}{\partial
v}\right\vert_{v=1} = \left. \frac{\partial f}{\partial y} \right\vert_T
\nmean_{r-1},
\ee
so that $\nmean_r=B^r$ with $B=\left. \frac{\partial f}{\partial y}
\right\vert_T$ .
Thus we still have an exponential decay of the average number of blobs
with distance and
\be
\label{eqn:fullB}
B= \frac{x T}{1-2xT} \left[ 1 -3xT + \frac{(1-xT)}{\ztwo} \left( 
3 x' \frac{\partial \ztwo}{\partial x'} + \sum_{j=1}^p 
2 \lambda_j' ( \lambda_j' -1 )
\frac{\partial \ztwo}{\partial \lambda_j'}  \right) \right]_{y=T},
\ee
where $\ztwo=\ztwo(x',\{\lambda'\})$.
One can also show that
\begin{eqnarray}
\left( \frac{\partial T}{\partial x} \right)_{\! \{\lambda\}}
 &=& \frac{1}{(1-B)}
\left. \left( \frac{\partial f(y)}{\partial x} \right)_{\! \{\lambda\}}
\right\vert_{y=T}  \\
 &=& \frac{T}{x(1-B)}  \left[B-x T + (1-xT)
\mtwo(x',\{\lambda'\}) \right] \label{eqn:dtdx} \\
 & \sim & (x_c -x)^{-\gst},
\end{eqnarray}
where
\be
\mtwo(x',\{\lambda'\}) \equiv \frac{x'}{\ztwo} \left( \frac{\partial
\ztwo}{\partial x'} \right)_{\! \{\lambda'\}}.
\ee
Thus as in the previous case, at $x_c$, $B_c=1$ in the tree phase and
on its boundary, where $\gst \ge 0$, and we have $0<B_c<1$ in the
magnetized phase. Using equations (\ref{eqn:fullB}) and (\ref{eqn:dtdx})
one can show, barring accidental cancellations, that $\gst=\half$ in
the tree phase and that on the boundary, $\gst^*=\gstwo /(\gstwo -1 )$,
which is just Durhuus' formula~\cite{Dur94}; where
$\gstwo(\{\lambda'\})$, which is negative, is the value of the string
susceptibility for $\ztwo(x',\{\lambda'\})$, that is,
\be
\left( \frac{ \partial^2 \ztwo}{\partial x'^2}\right)_{\!\{\lambda'\}}
\sim \left( x_c'(\{\lambda'\}) - x' \right)^{-(1 +
\gstwo(\{\lambda'\}))} .
\ee

\subsection{The spin-spin correlation function}
\label{sec:fullspin}

Next we consider the spin-spin correlation function. Rather than fixing
all the spins in the root blob, as we did for the toy model, it is
more convenient to add a set of $p$ external spins on the root, which
are all fixed to be $+1$. The partition function is as in
(\ref{eqn:general}), but with an extra factor for the link connecting
the external spins to the root blob, which we shall denote $W_{ext}$
(we will not add an extra factor of
$x$). Thus,
\be
\zone(x,\{\lambda\}) = \sum_{G \in \gone} x^N \frac{1}{2^{Np}}
\sum_{\{S\}} \left[ W_{ext} \prod_{<ij>} W_{ij} \right] ,
\ee
where the weight $W_{ij}$ for link $< \! ij \! >$ is given by the square
bracket in equation (\ref{eqn:general}). Note that when the sum over
spins is performed, the factor $W_{ext}$ just gives a contribution of
one, so that the function $\zone$ is unchanged by adding the external
spins. 
The average total spin at distance $r$ is
\be
\smean_r^+ = \frac{1}{\zone} \sum_{G \in \gone} x^N \frac{1}{2^{Np}}
\sum_{\{S\}} \left[ W_{ext} \left( \prod_{<ij>} W_{ij} \right)
 \sum_{k, \alpha} S_k^\alpha \right] ,
\ee
where the last summation is over all spins, $k$, at distance $r$ and
all spin flavours, $\alpha$.

\begin{figure}[tbh]
\caption[l]{Formula for $\gover_r$}
\label{fig:gr}
\begin{picture}(100,20)(0,0)
\centerline{\epsfbox{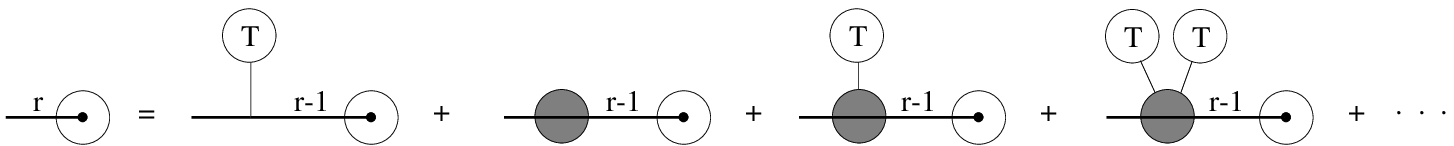}}
\end{picture}
\end{figure}

Consider $\zone \smean_r^+$, the only non-zero contributions come from
the configurations in which there is a flavour $\alpha$ line from the
external spins to vertex $k$, in addition to the usual sets of closed
loops. Define $\gover_r$ by
\be
\zone \smean_r^+ = p \sum_{G \in \gone} x^N \frac{1}{2^{Np}}
\sum_{\{S\}} \left[ W_{ext} \left( \prod_{<ij>} W_{ij} \right)
 \sum_{k} S_k^1 \right] = p \gover_r ,
\ee
Then $\gover_r$ is a sum over all rooted graphs and all loop
configurations that have a flavour~$1$ line from the external spins to
a vertex $k$ at distance $r$; the location of the vertex $k$ is also
summed over. However, we can write $\gover_r$ in terms of
$\gover_{r-1}$. This is shown diagrammatically in fig.~\ref{fig:gr},
where links included in the flavour $1$ line are drawn thicker than
those which are not included. The symbol on the left-hand side is used
to represent $\gover_r$, it has a flavour~$1$ line of length $r$, note
that we are still measuring distances from the root blob (not from the
external spins).

That is,
\be
\label{eqn:fullD}
\gover_r= \left[ 2 \lambda_1 x T + \ztwo_2(x',\{\lambda'\}) \lambda_1
(1-2xT) \right] \gover_{r-1},
\ee
where $ \ztwo_2(x,\{\lambda\})$, which is drawn as a shaded circle with a
thick line passing through it, is the partition function for 1PI
graphs with two distinct legs and a flavour $1$ line running through
(it does not include factors of $\lambda_1$ on the two legs); $x'$
and $\{\lambda'\}$ are given by (\ref{eqn:fullmap}) evaluated at
$y=T$. Define $D$ to be equal to the square bracket above, then
$\gover_r=D^r \gover_0$. So that we have an exponential decay with
distance,
\be
\label{eqn:fullmagdecay}
\smean_r^+ = \frac{p}{\zone} D^r \gover_0 = D^r \smean_0^+.
\ee
The total average spin is
\be
\smean^+=\smean^+_0 \frac{1}{1-D}.
\ee
 The magnetization in the grand canonical ensemble is given by 
\be
\label{eqn:fullmag}
{\cal M} = \frac{\smean^+}{p \Nmean} = \frac{\smean^+_0}{p \Nmean_0}
\left( \frac{1-B}{1-D} \right)
= {\cal M}_0 \left( \frac{1-B}{1-D} \right),
\ee
where ${\cal M}_0$ is the magnetization of the root blob ($0 \le {\cal
M}_0 \le 1$). We have used the fact that the average number of
vertices at distance $r$ is $\Nmean_r = B^r \Nmean_0$, so that $\Nmean
= \Nmean_0/(1-B)$. This is easily shown by defining
$G_r'=f^{(r)}(f(T)\vert_{xz})$ and following the derivation of
(\ref{eqn:nmeanr}). 
Note that, at $x_c$, in the tree phase $B_c=1$, so that ${\cal M}=0$
throughout this phase as one might expect. 

In the next section it is assumed that $D<1$, that is, that the
spin-spin correlation length does not diverge; the easiest way to
justify this would be to show that $D<B$, as we already know that $B\le 1$.
The following calculation will make this assumption plausible.
{}From (\ref{eqn:fullmagdecay}) one clearly has $D
\le B$, (at least provided that $\smean_0^+ \neq 0$), since
$\smean_r^+ \le p \Nmean_r$. To improve this inequality we need
an equation for $B$ which is simpler than (\ref{eqn:fullB}).
Define $G_r''$ to be the partition function for rooted graphs, with a
marked vertex at distance $r$ --- it is a sum over all graphs, $G \in
\gone$, all loop configurations and all ways of marking a vertex at
distance $r$; that is, $G_r''=T\Nmean_r$.
Then following the derivation of (\ref{eqn:fullD}) we
have
\be
G_r''=\left[ 2xT + \ztwo_t\xld (1-2xT) \right]
G_{r-1}''=B G_{r-1}'' ,
\ee
where $\ztwo_t$ is the partition function for 1PI graphs, with two
distinct legs --- it is a sum over all such graphs and all loop
configurations. Note, $\ztwo_t$ differs from $\ztwo_2$ in that there
is no flavour~1 line running through it.
Thus,
\begin{eqnarray}
B &=& \phantom{\lambda_1} \ \ 2xT + \ztwo_t\xld (1-2xT)  \\
D &=& \lambda_1 \left[2xT + \ztwo_2\xld (1-2xT) \right] .
\end{eqnarray}
Now, if one could show that $\ztwo_t\xld \ge \ztwo_2\xld$, then this
would imply that $D \le \lambda_1 B$. 

\begin{figure}[bth]
\caption[l]{Tentative phase diagram}
\label{fig:phase}
\begin{picture}(100,77)(0,0)
\centerline{\epsfbox{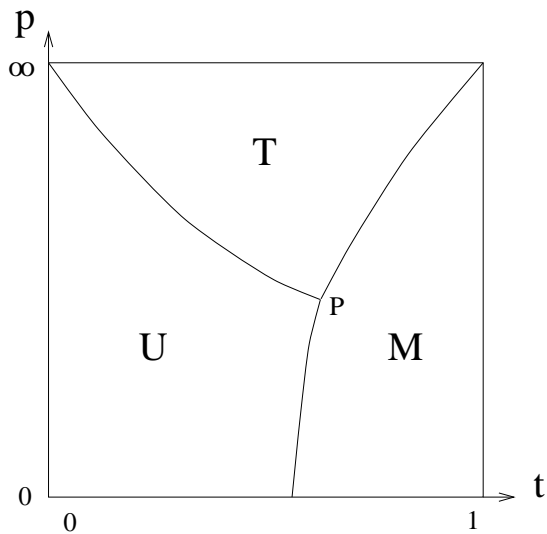}}
\end{picture}
\end{figure}

Consider $\ztwo_{+-}\xld$, the
partition function for 1PI graphs with two distinct legs, in which the
flavour~1 spins on the two vertices connected to the legs are held
fixed at $+1$ and $-1$ respectively; all other spins are summed
over. Then one can easily see that $\ztwo_{+-}\xld = \ztwo_t\xld -
\ztwo_2\xld$, since the only non-vanishing terms correspond either to sets of
closed loops, or configurations in which there is also a flavour~1
line running between the two fixed spins; this last set of terms is
multiplied by the product of the fixed spins, namely $-1$.
For small enough values of $\{\lambda'\}$, $\ztwo_{+-}\xld$ is
manifestly positive; the weights $W_{ij}$ for each link will be positive
for any spin configuration (see equation (\ref{eqn:general})), and hence
$\ztwo_{+-}$ is just the sum of positive terms. At $\lambda_j'=1$ for all $j$,
the system is completely magnetized and hence $\ztwo_{+-}=0$. One
would expect that $\ztwo_{+-}\xld >0$ for other values of
$\{\lambda'\}$. In which case $D \le \lambda_1 B$, with $D<B$ in
general and $D=B$ only when $\lambda_j=\lambda_j'=1$ for all $j$.
Thus in general $D<1$ and the spin-spin correlation length does not
diverge. Note that our definition of distance is only good in the
tree-like phase and on its boundary, and hence this result does not
say anything about the possible divergence of the proper spin-spin
correlation length (defined using the geodesic distance) at the
magnetization transition, between the U and M phases
(see figure~\ref{fig:phase}).

Figure~\ref{fig:phase} shows a tentative phase diagram for the
standard model defined by (\ref{eqn:full}); T is the
(unmagnetized) tree-like or branched polymer phase,
U the unmagnetized phase and M the magnetized
phase. See~\cite{paper3,paper2} for a discussion of various possible
alternative phase diagrams. Each line of this phase diagram for
constant $p$ (where $p$ is a non-negative integer) corresponds to a
line through the $p$-dimensional phase space of the generalized model.
Within the U and M phases the model behaves in a similar fashion to
the pure gravity case and it seems probable that $\gst=-\half$
throughout both these phases. The transition between the
magnetized (M) and unmagnetized (U) phases is caused by ${\cal M}_0$
vanishing as the coupling constants $\{\lambda\}$ are
varied. From the KPZ results, we expect to have $-\half  \le \gst^* \le 0$
along the corresponding critical line, at least for $p\le 2$. 
The tree-like phase has $\gst=\half$ throughout, the generic
value for branched polymers.

\subsection{Critical exponents}
Now consider the transition from the magnetized to the tree phase,
which is caused by $B_c \to 1$. The behaviour of the full model at this
transition is very similar to that of the toy model. Following the
analysis in section~\ref{sec:toycrit}, we define a mass by $m=-\ln
B$. As before there is an exponential decay, so we will take $\eta=1$,
which gives $\gst=\nu$ provided that the scaling relation
$\gst=\nu(2-\eta)$ holds. Consider the magnetic exponents, at $x_c$
and fixed integer $p$ we vary the $\{\lambda\}$ in order to approach the phase
transition; the variable $\Delta \lambda$ will be used to parameterize
this, with $\Delta \lambda=0$ at the transition. Then by definition
$m \sim (\Delta \lambda)^{\nu_m}$, but $m=-\ln B_c$ so that $B_c \sim 1 -
(\Delta \lambda)^{\nu_m}$. Now $\beta_m$ is defined by ${\cal M} \sim
(\Delta \lambda)^{\beta_m}$, but ${\cal M} \sim 1-B_c$ from
(\ref{eqn:fullmag}), thus $\beta_m=\nu_m$; note, we have assumed
that we are not close to any point at which ${\cal M}_0$ vanishes
($P$ in fig.~\ref{fig:phase} is such a point). 
It has also been assumed that $D_c \neq 1$ at the transition, i.e.\ 
that the spin-spin correlation length does not diverge; this was
justified in the previous section.
If we assume that
$\gamma_m=\nu_m(2-\eta_m)$ with $\eta_m=1$, then we have
$\gamma_m=\nu_m=\beta_m$. Assuming the scaling relations
$2-\alpha=\nu_m d_H^*$ and $\alpha + 2 \beta_m + \gamma_m=2$, gives
$d_H^*=3$. The calculation in~\cite{AmbWat95}, which gave that $\nu
d_H=1$, still applies, so that $\nu^*=\third$ and hence
$\gst^*=\third$. This is encouraging as the analysis in~\cite{Dur94}
leads one to suppose that $\gst^*=\third$ provided $\gst=-\half$
in the magnetized phase. Also, applying the relation $\beta_m \delta_m
= \beta_m + \gamma_m$ gives $\delta_m=2$.

Now $B$ is given by (\ref{eqn:fullB}) and in the magnetized phase at
$x_c$, $x'=\xctwo(\{\lambda'\})$. Provided that we are away from points
at which ${\cal M}_0$ vanishes, $\ztwo(x',\{\lambda'\})$ should be
analytic and Taylor expandable about $\{\lambda'^*\}$ (the value of
$\{\lambda'\}$  at the M to T phase transition). Assuming that
$x_c(\{\lambda\}) T_c(\{\lambda\})$ is expandable in powers of $\Delta
\lambda$ as far as $\Delta \lambda$, then barring accidental
cancellations $B_c \sim 1 - \Delta \lambda$ and hence $\nu_m=
\beta_m=\gamma_m=1$ and $\alpha=-1$. 

Thus on the assumption of various
scaling relations and that our definitions of $m$ and ${\cal M}$ are
adequate,
and barring problems such as accidental cancellations, the
set of exponents given in table~\ref{tab:exp} should also apply for
the full model for the M to T transition, away from point $P$. 
Again it is not clear what the value of $d_H$ is, within the M phase;
however one might guess that $d_H=4$, the pure gravity
value~\cite{AmbWat95,CTBJ,AJW},
throughout the U and M phases (in which case it is probable that
$\nu=1/4$ and $\eta=4$ here). Note that the mechanism by which
the U to T transition occurs is
essentially the same and so we would expect to have the same set of
exponents, although the magnetic exponents are not relevant since
${\cal M}=0$ in both the U and T phases. Perhaps one should note that
a low-temperature expansion matrix model calculation in the limit $c
\to \infty$~\cite{Wexler} gives that $\gst^*=\third$ at the transition
between the tree and magnetized phases, and that the truncated model
studied in~\cite{paper3}, which is supposed to approximate the U to T
transition, has $\gst^*=\third$ and $\alpha = -1$, agreeing with our
predictions. In addition the q-state Potts model at large
$q$~\cite{Wexlerq} , which is equivalent to the multiple Ising model
in the limit $q=p=\infty$ and may be related to it for finite $q$, has
a branched polymer region (with $\gst=\half$) separating two pure
gravity regions (with $\gst=-\half$) and has $\gst^*=\third$  and
$\alpha=-1$ at both transitions.

At point $P$, the above
argument fails since ${\cal M}_0 \to 0$ and hence it is possible to
have $\beta_m \neq \nu_m$; some of the other assumptions may break
down here too. Thus at this point one might expect to get
a different set of exponents. In any case the calculations in this
section only apply for non-negative integer $p$ and it is not entirely
clear whether $p^*$, the value of $p$ at point $P$, is an integer.
It is tempting to suppose that $p^*=2$, in which case one could
understand the breakdown of the KPZ formula in terms of the appearance
of the branched polymer phase, however Monte Carlo simulations do not
seem to support this hypothesis~\cite{MC}.

\section{Conclusion}
\label{sec:concl}
In this article we have studied correlation functions in
two-dimensional quantum gravity coupled to Ising spins.
For the toy model
two approximations are used: (i) the spin configurations allowed
are the ones dominant in the low temperature expansion of
multiple Ising spins on dynamically triangulated surfaces
and (ii) the distance between two spins is identified
with the number of links separating 1PI subgraphs
along a path connecting the spins. In the full model only the second
approximation is made.
We do not expect
this last approximation to be important for values
of $\beta$ where $\gamma_{str}(\beta) >0$, i.e.\ for values
of $\beta$ where the average number of vertices of a generic
$\phi^3$ graph diverges for $x \mbox{{\scriptsize $\nearrow$}}
\,x_c(\beta)$, since
the average number of vertices in an 1PI part of a generic
$\phi^3$ graph in our ensemble is very small. However,
if $\gamma_{str}(\beta) < 0$ the average number of vertices
in a generic $\phi^3$ graph is itself small, even at the
critical point $x_c(\beta)$ and our definition
of distance can not be used to extract in a reliable way
the fractal properties of the ensemble of $\phi^3$ graphs.
Approximation (i) has been shown
to be exact in the $c\to \infty$ limit~\cite{Wexler} and numerical
simulations~\cite{MC} 
suggest that it is an excellent approximation except for the smallest values
of $\beta$ if  many Ising spins are coupled to two-dimensional
gravity. For moderate values of $c$ it is difficult to
judge if approximation (i) is reliable all the way down to
$\beta^*$. For $c=1/2$ and $c=1$ (i) not a good approximation
for $\beta \in [0,\beta^*]$.

The toy model allows us to analyze
the two-point function (the puncture-puncture correlator)
and the spin-spin correlator as a function of the
distance $r$. For $\beta \leq \beta^*$  we
have $\gamma_{str} > 0 $ and the fractal structure is
reliably extracted from the two-point function.
We find that $d_H =2$ for $\beta < \beta^*$, while it jumps to
$d_H^* =3$ at $\beta^*$. Whilst we can not calculate $d_H$ for
$\beta>\beta^*$, one might expect it to equal four, the pure gravity
value~\cite{AmbWat95}. The fact that $d_H$
is different from 4 for $\beta \leq \beta^*$ is an indication of the very
strong interaction between gravity and matter for $c > 1$.

A further indication of the strong link between geometry and
matter configurations present in the model is found in the scaling
relations for the magnetic exponents. They are found from the
behaviour of the {\it two-point function} $\langle  n \rangle_r$
in the region $\beta \geq \beta^*$. The exponential decay of
$\langle  n \rangle_r$ {\it at} $x_c (\beta)$
as a function of the distance $r$
defines a length scale
\be
\xi(\beta)  \sim \frac{1}{\beta-\beta^*},
\ee
which diverges for $\beta \to \beta^*$ and allows us to define
the magnetic exponents using the standard hyper-scaling relations
(note that $\xi(\beta) = \infty$ for $\beta \le \beta^*$).
The same exponents are found, for the toy model,
by a direct calculation in the
canonical and grand canonical ensembles. 
Strictly speaking, it seems more natural to define
magnetic scaling properties from one of the spin-spin correlators,
i.e.\ either $\langle \Sigma S\rangle_r^{++\cdots+}$ or ${\cal M}_r$,
however the corresponding correlation length does not diverge at
$\beta^*$. This is an indication of the geometric nature of the
transition. 

For the full model, a summation over all spin configurations is
performed and the only approximation made is in the measure of
distance that is used. As before there is an exponential decay of both
the two-point function and the spin-spin correlation function,
and one can define, in the magnetized phase, a
geometric correlation length, which diverges as the tree phase is
approached. Based on the analysis in section~\ref{sec:fullspin} we do
not expect the spin-spin correlation length, at least as we have
defined it, to diverge at the magnetized to tree-like transition.
Using the geometric length scale and the scaling relations, which have been
shown to hold for the toy model, all the critical exponents are
calculated for the branched polymer phase and its boundary, on the
basis of a small number of assumptions. The exponents are the same as
those for the toy model (see table~\ref{tab:exp}) and again $d_H=2$
within the tree phase and $d_H$ is equal to three on the critical line,
showing the strong interaction between the matter fields and the
geometry.

Some interesting questions remain to be answered, such as what happens
outside the branched polymer phase, in particular what the values of
the geometric exponents and $d_H$ are, and whether there is a diverging
length associated with the spin-spin correlation function at the
magnetization transition (between the U and M phases). Unfortunately
our measure of distance is inadequate in this region and it will
require further work using probably the full geodesic distance in
order to determine this.
It would also be useful to locate point $P$ on the phase diagram, as
it has been suggested~\cite{AJW,MC} that $p^*>2$, in which case there is an
intermediate region, for $2<p<p^*$, between the KPZ regime and the
branched polymer phase.

Finally, it should perhaps be noted that most of the considerations in
this paper also apply to the O(n) models~\cite{onmodels,charlotte}, as
these are just a special case of our generalized multiple Ising model.

\subsection*{Acknowledgements}
MGH would like to acknowledge the support of the Royal Society through
their European Science Exchange Programme.


\end{document}